\documentclass[useAMS,usenatbib,referee]{biom}
\usepackage{amsmath}
\usepackage{graphicx}
\usepackage{url}

\def\bSig\mathbf{\Sigma}

\title[Two-Stage Penalized Regression Screening]{Two-Stage Penalized Regression Screening to Detect Biomarker-Treatment Interactions in Randomized Clinical Trials}

\author
{Jixiong Wang$^{1, *}$\email{jixiong.wang@mrc-bsu.cam.ac.uk}, Ashish Patel$^{1, **}$\email{ashish.patel@mrc-bsu.cam.ac.uk}, James M.S. Wason$^{1, 2, ***}$\email{james.wason@mrc-bsu.cam.ac.uk}, and Paul J. Newcombe$^{1, ****}$\email{paul.newcombe@mrc-bsu.cam.ac.uk} \\
$^1$MRC Biostatistics Unit, University of Cambridge, Cambridge CB2 0SR, U.K. \\
$^2$Population Health Sciences Institute, Newcastle University, Newcastle upon Tyne NE2 4BN, U.K.}

\begin{document}


\date{{\it Received March} 2020. {\it Revised November} 2020.  {\it
Accepted December} 2020.}



\pagerange{\pageref{firstpage}--\pageref{lastpage}} 
\pubyear{2021}
\artmonth{January}


\doi{10.1111/biom.13424}


\label{firstpage}


\begin{abstract}
	High-dimensional biomarkers such as genomics are increasingly being measured in randomized clinical trials. Consequently, there is a growing interest in developing methods that improve the power to detect biomarker-treatment interactions. We adapt recently proposed two-stage interaction detecting procedures in the setting of randomized clinical trials. We also propose a new stage 1 multivariate screening strategy using ridge regression to account for correlations among biomarkers. For this multivariate screening, we prove the asymptotic between-stage independence, required for family-wise error rate control, under biomarker-treatment independence. Simulation results show that in various scenarios, the ridge regression screening procedure can provide substantially greater power than the traditional one-biomarker-at-a-time screening procedure in highly correlated data. We also exemplify our approach in two real clinical trial data applications.
\end{abstract}

%



\maketitle


%

\section{Introduction}

Recent developments in medicine have seen a shift toward targeted therapeutics. It has been shown that individual variability can often contribute to differences in response to the same treatment. For example, patients with leukemia respond to the treatment with all-trans retinoic acid if they have the PML-RARA translocation \citep{sawyers2008cancer}. Conversely, use of some drugs can lead to increased risk to patients with specific genetic variants, e.g. the Class II allele HLA-DRB1*07:01 has been associated with lapatinib-induced liver injury \citep{parham2016comprehensive}. Detecting such interactions between biomarkers and treatments in randomized clinical trials is of growing interest.

Discovering biomarker-treatment interactions helps identify predictive biomarkers: biomarkers which influence treatment efficacy can be used to find the subgroup of patients who are most likely to benefit from the new treatment, as well as to predict subgroup treatment effects. Consequently, new adaptive design approaches can be used in settings where there are genetically-driven subgroups to improve efficiency \citep{wason2015bayesian}. Furthermore, the discovery of novel biomarker-treatment interactions may result in the identification of new disease susceptibility loci, providing insights into the biology of diseases. Such outcomes are very much aligned with the goals of precision medicine: to enable the provision of ``the right drug at the right dose to the right patient'' \citep{collins2015new}.

Detecting biomarker-treatment interactions in large-scale studies of human populations is a non-trivial task, which faces several challenging problems \citep{mcallister2017current}. Traditional interaction analysis, using regression models to test biomarker-treatment interactions one biomarker at a time, may suffer from poor power when there is a large multiple testing burden, for example when performing such analysis on a genome-wide scale for genetic biomarkers. Standard genotyping microarrays measure half a million or more variants and, when combined with whole genome imputation, can lead to millions of biomarkers to consider. Another type of omics, metabolomics - the measurement of metabolite concentrations in the body - may have a more direct effect on drug efficacy and is also becoming increasingly widely assayed \citep{beckonert2007metabolic}.

In the context of gene-environment interaction studies, there is now a significant literature of statistical methods, which exploit aspects of the study design to improve power thus mitigating the multiple testing burden. These include case-only tests \citep{piegorsch1994non}, empirical Bayes \citep{mukherjee2008exploiting}, Bayesian model averaging \citep{li2008detecting}, and two-stage tests with different screening procedures \citep{kooperberg2008increasing, murcray2008gene, gauderman2013finding, wason2012general}. To alleviate the multiple testing burden, two-stage methods use independent information from the data to perform a screening test to select a subset of genetic biomarkers, and then only test interactions within this reduced set. Since there is a clear analogy to gene-environment interaction problems, in this paper, we will examine how existing gene-environment interaction testing methods may be modified so that they are transferable to the biomarker-treatment setting \citep{dai2009semiparametric, dai2016augmented, wang2016twophaseind}. One significant drawback of the traditional two-stage approach testing each biomarker one at a time is that the univariate screening tests will harm power of the overall two-stage procedure when there exist substantial correlations between biomarkers. We also propose a novel screening test in this two-stage framework, which utilizes ridge regression to model correlated high-dimensional data at stage 1. We prove that this new two-stage method is able to preserve the overall family-wise error rate given independence between the treatment and biomarkers. Furthermore, it is shown by simulations and real data applications that the new method can provide better performance than traditional one-biomarker-at-a-time approaches for correlated biomarkers. In the context of more general variable selection settings, screening strategies have been explored to focus algorithms on a reduced search space \citep{fan2008sure, wang2016high}. In this work, we explore the use of variable pre-screening specifically to help identify interactions and the condition required for controlling the family-wise error rate.

\section{Methods}
\label{sec:methods}

\subsection{Standard Single-Step One-Biomarker-at-a-Time Interaction Tests}

In the context of randomized clinical trials, one can test each biomarker in turn for a biomarker-treatment interaction using the following linear model 
\begin{eqnarray}
E(Y_i \mid X_{ij}, T_i) = \beta_{0_j} + \beta_{X_j} X_{ij} + \beta_T T_i + \beta_{X_j \times T} X_{ij} \times T_i \label{eqn:intStandard}
\end{eqnarray}
with $Y_i$ denoting the response outcome, $T_i$ the binary treatment-control indicator, and $X_{i1}, \dots, X_{im}$ representing the values of $m$ biomarkers, for the $i$th patient. The null hypothesis $\beta_{X_j \times T} = 0$ could be tested for each $j = 1, \dots, m$, e.g. using a Wald test with the Bonferroni correction applied to preserve the family-wise error rate.

The number of biomarkers $m$ to be considered is potentially large. Given the desired overall family-wise error rate $\overline{\alpha}$, a Bonferroni correction \citep{dunn1961multiple} requires an adjusted significance level for each individual test to be $\overline{\alpha} / m$. Although the Bonferroni correction is typically used for its simplicity and flexibility, with regard to our interest in high-dimensional interaction testing it is worth exploring whether other procedures are able to provide improved efficiency. In Web Appendix~A, we demonstrate theoretically some alternative family-wise error rate controlling methods \citep{vsidak1967rectangular, holm1979simple} can only provide a small improvement across the settings we consider in this paper: when $m$ is large and only a small subset of biomarkers have true interactions with treatment.

\subsection{Two-Stage Interaction Tests with Some Existing Screening Methods}
\label{sec:screeningTests}

Two-stage approaches use a screening test as a filtering stage (stage 1) to select a subset of biomarkers, and then in stage 2, only test interactions within the reduced set of biomarkers, thus increasing power. To preserve the overall family-wise error rate, two-stage approaches rely on the stage 1 screening tests being independent of the final stage 2 tests.

A common stage 1 screening test used in two-stage interaction testing is a marginal association test \citep{kooperberg2008increasing}. Considering this type of screening test in the clinical trial setting, the marginal effect of a biomarker on the outcome can be measured in a regression model of the form 
\begin{eqnarray}
E(Y_i \mid X_{ij}) = \delta_{0_j} + \delta_{X_j} X_{ij} \label{eqn:marginalScreening}
\end{eqnarray}
The screening procedure is conducted by testing the null hypothesis $\delta_{X_j} = 0$ for $j = 1, \dots, m$, with a pre-specified significance level $\alpha_1 \in (0, 1)$. In stage 2, one then tests interactions using the one-biomarker-at-a-time model \eqref{eqn:intStandard} within the set of biomarkers selected at stage 1. Another way to utilize stage 1 information is to test all $m$ biomarkers in stage 2 using weighted significance levels, that add up to the targeted error rate $\overline{\alpha}$, based on ordered biomarkers from stage 1. One possible weighting scheme \citep{ionita2007genomewide} is: the $B$ most significant biomarkers, i.e. with lowest $p$-values in stage 1, are compared with an adjusted significance level $(\overline{\alpha} / 2) / B$, the next $2B$ biomarkers are compared with $(\overline{\alpha} / 4) / (2B)$, ..., the next $2^k B$ biomarkers are compared with $(\overline{\alpha} / 2^{k + 1} ) / (2^k B)$, and so on.

The motivation of conducting marginal association tests to screen for candidate interaction tests is that we expect a biomarker that has an interaction with the treatment for the disease will also show some level of marginal association with the response. However, it is also possible that the biomarker's main association with response and the interaction effect may be in opposite directions. When this is the case, a marginal screening strategy would downgrade due to the first stage test statistic having low power.

To preserve the overall family-wise error rate, a key requirement to apply the two-stage approach is the independence between stage 1 and 2 tests. Both \citet{murcray2008gene} and \citet{dai2012two} proved that: with stage 1 and 2 test statistics being asymptotically independent and $m^*$ defined as the number of stage 1 selected biomarkers, using a Bonferroni adjusted significance level $\alpha = \overline{\alpha} / m^*$ at stage 2 to test interactions within the reduced set is sufficient to preserve the overall family-wise error rate of the two-stage procedure under $\overline{\alpha}$.

In the context of gene-environment interaction studies, an alternative type of screening is testing the correlation between a gene and the environmental factor \citep{murcray2008gene}. This type of screening requires case-control sampling for a rare response endpoint, thus it can be useful for detecting biomarker-treatment interactions in large prevention trials. However, such a screening procedure is not generally applicable in randomized clinical trials, where the rare response condition does not hold. In this case, the trial population represents the entire dataset and cases (responders) are not ``oversampled". We make this argument and also discuss the applicability of other related proposals more formally in Web Appendix~B.

\subsection{New Stage 1 Penalized Regression Screening Procedure Accounting for Biomarker-Biomarker Correlations}

One drawback of existing two-stage interaction testing procedures is that biomarkers are only screened one at a time in stage 1. This ignores correlations between the biomarkers. In a high-dimensional, low-sample-size data set, an ordinary multivariate regression analysis testing each predictor, while accounting for correlations with the other predictors, is not feasible.  Therefore we considered penalized regression methods to model correlated high-dimensional data. These techniques have improved the development of risk predictors from high-dimensional genomic information \citep{wu2009genome, newcombe2017weibull}.

We propose a new stage 1 multivariate screening test of the following form to account for biomarker-biomarker correlations 
\begin{eqnarray}
E(Y_i \mid X_{i1}, \dots, X_{im}) = \delta_0 + \delta_T T_i + \sum_{j = 1}^m \delta_{X_j} X_{ij} \label{eqn:ridgeScreening}
\end{eqnarray}
This multivariate version of the marginal association screening test also includes the treatment main effect term. This is necessary to preserve the independence between stage 1 screening and stage 2 interaction tests as described later.

To fit this multivariate model, we use ridge regression, which applies regularization to avoid overfitting in high-dimensional low-sample-size problems. Typically, the objective of ridge regression is to minimize a loss function $L_n$ along with an $L_2$ regularization term: $L_n(\boldsymbol{\delta}) + \lambda_n ||\boldsymbol{\delta}||_2^2$, where $||\boldsymbol{\delta}||_2^2 = \delta_T^2 + \sum_{j = 1}^{m} \delta_{X_j}^2$ and $\lambda_n$ is the regularization parameter. Ridge shrinks all the estimated coefficients towards zero, but will not set them exactly to zero. For use in a two-stage interaction testing strategy, we propose ordering the biomarkers based on the ridge coefficients obtained from stage 1, and then use the resulting ranking to determine varying significance thresholds across buckets of markers during stage 2 one-at-a-time interaction tests according to the weighting scheme described in Section~\ref{sec:screeningTests}.

\subsection{Proof of Independence between Stage 1 Penalized Regression Screening and Stage 2 Standard Interaction Tests}

In this section, we show that independence between stage 1 and stage 2 test statistics holds for stage 1 ridge regression screening tests.

For the $i$th subject, let $Y_i$ denote the outcome variable, $\boldsymbol{X}_i = (T_i, X_{i1}, \dots, X_{im})^T$ be a vector of the binary treatment-control indicator and $m$ biomarkers. Consider the proposed stage 1 marginal association screening test based on the multivariate model of the form
\begin{eqnarray}
E(Y_i \mid \boldsymbol{X}_i) = \boldsymbol{X}_i^T \boldsymbol{\delta} \nonumber
\end{eqnarray}
where $\boldsymbol{\delta} = (\delta_T, \delta_{X_1}, \dots, \delta_{X_m})^T$. The model underlying the stage 2 standard one-biomarker-at-a-time interaction test is of the form
\begin{eqnarray}
E(Y_i \mid \boldsymbol{V}_{ij}) = \boldsymbol{V}_{ij}^T \boldsymbol{{\beta}}_j\quad (j = 1, \dots, m) \nonumber
\end{eqnarray}
where $\boldsymbol{V}_{ij} = (X_{ij}, T_i, X_{ij} T_i)^T$ and $\boldsymbol{\beta}_j = (\beta_{X_j}, \beta_{T_j}, \beta_{X_j \times T})^T$. The above forms ignore intercepts without loss of generality. Homogeneity of variance is assumed, i.e. $var(Y_i \mid \boldsymbol{X}_i)$ and $var(Y_i \mid \boldsymbol{V}_{ij})$ are assumed to be constants. We first show the between-stage asymptotic independence for the stage 1 multivariate regression marginal association estimator without regularization.
\begin{theorem}
\label{theorem1}
For any $j = 1, \dots, m$, if $X_{ij}$ is independent of $T_i$, and, $E(T_i) = 0$ or $E(X_{ij}) = 0$ (i.e. $T_i$ or $X_{ij}$ is centered around $0$), then under the null hypothesis $\beta_{X_j \times T} = 0$, 
\begin{eqnarray}
cov\{n^{1 / 2} (\widehat{\delta}_{X_j}^0 - \delta_{X_j}), n^{1 / 2} (\widehat{\beta}_{X_j \times T} - \beta_{X_j \times T})\} \rightarrow 0 \nonumber
\end{eqnarray}
in probability, where $\widehat{\delta}_{X_j}^0$ and $\widehat{\beta}_{X_j \times T}$ are the maximum likelihood estimators for unknown parameters $\delta_{X_j}$ and $\beta_{X_j \times T}$ respectively without regularization (i.e. $\lambda_n = 0$).
\end{theorem}

The proof is provided in the appendix. Previous works \citep{dai2012two} have demonstrated that the stage 1 univariate marginal association screening tests are independent with the stage 2 one-biomarker-at-a-time interaction tests. Theorem~\ref{theorem1} extends this to show independence still holds when stage 1 tests are extended to a multivariate regression. Our proof relies on: 1) the inclusion of the treatment main effect in the multivariate regression of the form \eqref{eqn:ridgeScreening}; 2) an assumption of independence between the treatment assignment and biomarker values, which is valid in randomized clinical trials. The proof in \citet{dai2012two} for the univariate marginal association screening tests is more general; it does not depend on biomarker-environment independence and it also holds for generalized linear models.

Next we establish the asymptotic distribution of the ridge estimator.
\begin{lemma}
\label{lemma1}
Under standard regularity conditions \citep[p. 51-52]{van2000asymptotic} and if $\lambda_n = O(n^{1 / 2})$, i.e. $\lim_{n \rightarrow \infty} \lambda_n / n^{1 / 2} = \lambda_0 \ge 0$, then 
\begin{eqnarray}
n^{1 / 2} (\widehat{\boldsymbol{\delta}}^{\lambda} - \boldsymbol{\delta}) \rightarrow {\cal N}(- 2 \lambda_0 \boldsymbol{\Sigma}^{-1} \boldsymbol{\delta}, \sigma^2 \boldsymbol{\Sigma}^{-1}) \nonumber
\end{eqnarray}
in distribution, where $\widehat{\boldsymbol{\delta}}^{\lambda}$ is the ridge estimator, $\cal N$ is a normal distribution, $\sigma$ and $\Sigma$ are a constant and an invertible constant matrix.
\end{lemma}

Based on the asymptotic results derived in Lemma~\ref{lemma1} and Theorem~\ref{theorem1}, we are able to prove the asymptotic independence between the stage 1 ridge marginal association screening estimator and the stage 2 one-at-a-time interaction estimator in the following corollary.
\begin{corollary}
\label{corollary1}
For any $j = 1, \dots, m$, if $X_{ij}$ is independent of $T_i$, and, $E(T_i) = 0$ or $E(X_{ij}) = 0$ (i.e. $T_i$ or $X_{ij}$ is centered around $0$), then under the null hypothesis $\beta_{X_j \times T} = 0$, 
\begin{eqnarray}
cov\{n^{1 / 2} (\widehat{\delta}_{X_j}^{\lambda} - \delta_{X_j}), n^{1 / 2} (\widehat{\beta}_{X_j \times T} - \beta_{X_j \times T})\} \rightarrow 0 \nonumber
\end{eqnarray}
in probability, where $\widehat{\delta}_{X_j}^{\lambda}$ is the maximum likelihood estimator with the ridge penalty.
\end{corollary}
Proofs of Lemma~\ref{lemma1} and Corollary~\ref{corollary1} are given in Web Appendices~C and D.

\section{Results}
\label{sec:res}

\subsection{Simulation Study}

To evaluate performance of our proposed biomarker-treatment interaction testing procedure described above, we generated simulated data sets, each having $m = 1,000$ biomarkers. Data were simulated under the model $Y_i = \beta_0 + \beta_T T_i + \sum_{j = 1}^m (\beta_{X_j} X_{ij} + \beta_{X_j \times T} X_{ij} \times T_i) + \varepsilon_i$, where the treatment main effect was set to $\beta_T = 0.5$ and the intercept $\beta_0 = 0$. We partitioned the $1,000$ biomarkers into $50$ clusters of correlated biomarkers, containing $20$ biomarkers each. We denote the clusters $C_1 = \{X_1, \dots, X_{20}\}$, $C_2 = \{X_{21}, \dots, X_{40}\}$, and so on. One biomarker in the first cluster was ascribed a main effect and an interaction effect, i.e. $\beta_{X_1} = 0.5$ and $\beta_{X_1 \times T} = 1$. Four other biomarkers in four other different clusters were ascribed main effects on the trait without interactions, i.e. $\beta_{X_{21}} = \beta_{X_{41}} = \beta_{X_{61}} = \beta_{X_{81}} = 1.5$. All other biomarkers do not have direct effects on the outcome. Each biomarker $X_j$ was generated from a standard normal distribution ${\cal N}(0, 1)$ and the binary treatment assignment was drawn from a $Bernoulli(0.5)$ distribution, while $\varepsilon_i$ was generated from a normal distribution with standard deviation $5$. In this case, the proportion of variance explained by the true model is $0.292$. We consider two types of correlation patterns among biomarkers: 1) The $20$ biomarkers within each cluster are correlated with each other ($\rho = 0.6$), but there are no correlations between biomarkers in different clusters; 2) All biomarkers are independent of one another ($\rho = 0$). For each scenario, $1,000$ replicate data sets were generated to estimate power and family-wise error rates. Power for all the approaches is defined according to the idea of ``cluster discoveries'' in \citet{brzyski2017controlling} as $pr(\text{reject at least one } H_0^j \text{ for any } X_j \in C_i \mid \text{at least one } H_1^k \text{ is true for any } X_k \in C_i)$, where $H_0^j$ is the null hypothesis for $X_j$ and $H_1^k$ is the alternative hypothesis for $X_k$.

Four different screening procedures are compared: 1) ``Univariate screening (threshold)'': A selection of biomarkers to take forward to stage 2 is based on significance in a regression of response on the biomarkers one at a time, of the form \eqref{eqn:marginalScreening}. A significance level $\alpha_1 = 0.05$ is used without adjustment for each stage 1 biomarker test. 2) ``Univariate screening (rank)'': All biomarkers are taken forward to stage 2, and the stage 1 $p$-value ranking is used to conduct a stage 2 weighted hypothesis test described in Section~\ref{sec:screeningTests} with $B = 5$ \{a number recommended by \cite{gauderman2013finding}\}. 3) ``Ridge screening (rank)'': Ridge regression is used to estimate marginal effects at stage 1. Then all biomarkers are ordered based on these stage 1 coefficients and the rank will be used by the stage 2 weighted hypothesis test with $B = 5$. The optimal $\lambda_n$ is chosen based on $5$-fold cross-validation errors. The R package \textbf{glmnet} \citep{friedman2010regularization} was used. 4) ``No screening'': A standard single-step interaction test of the form \eqref{eqn:intStandard}, targeting an overall family-wise error rate $\overline{\alpha} = 0.05$, is performed as a baseline comparator (with a Bonferroni correction applied with $m = 1,000$) and also as the stage 2 test for all three two-stage approaches described above.
\begin{figure}
\includegraphics[width=1\linewidth]{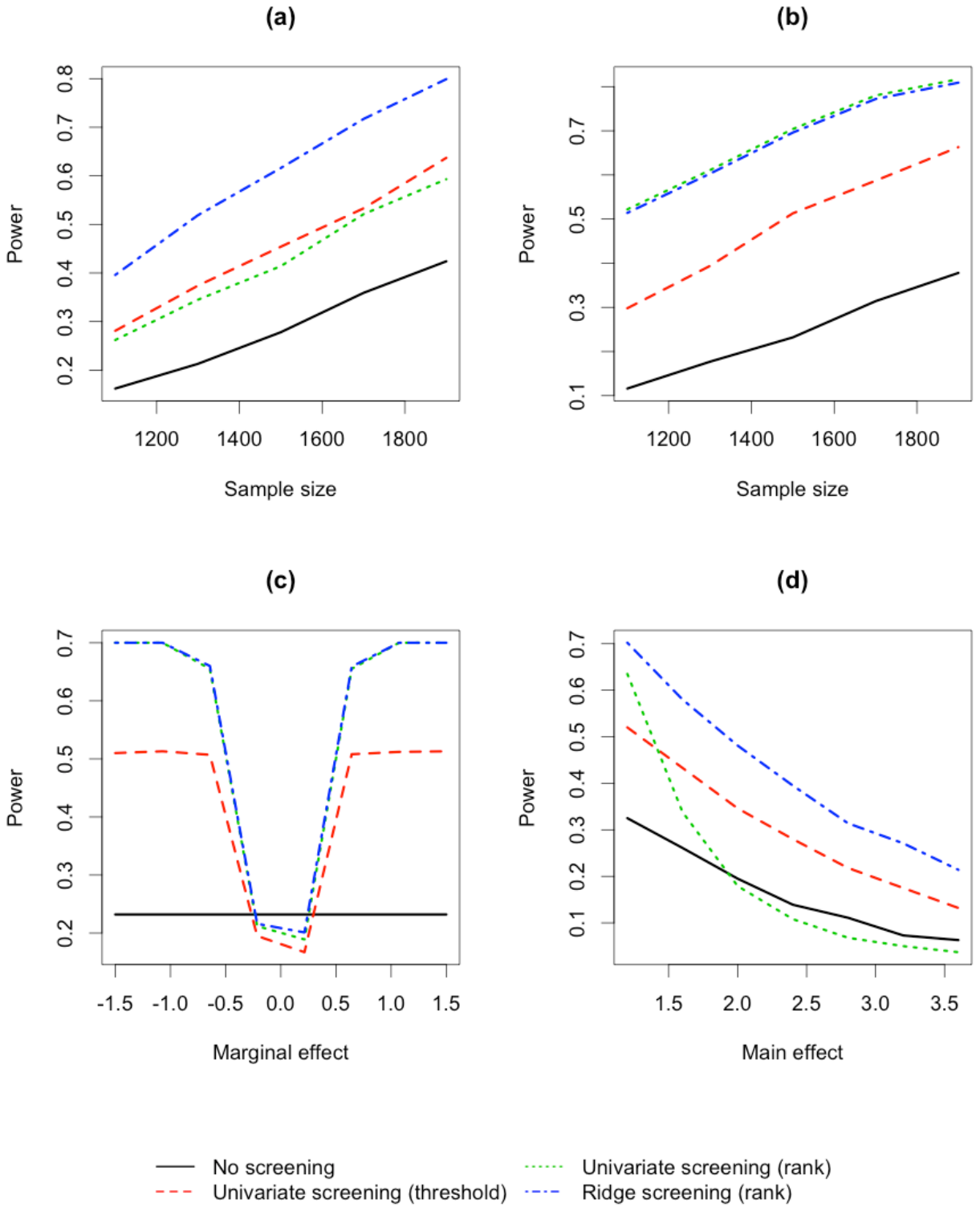}
\caption{Comparison of two-stage interaction tests with different screening testing procedures. Four were compared: univariate screening (threshold) (long dashes), univariate screening (rank) (short dashes), ridge screening (rank) (dot-dash), and no screening (solid). The four panels represent: (a) highly correlated biomarkers ($\rho = 0.6$), (b) independent biomarkers ($\rho = 0$), (c) independent biomarkers ($\rho = 0$, sample size of $1,500$), changing the main effect of the interacting biomarker $\beta_{X_1}$, (d) highly correlated biomarkers ($\rho = 0.6$, sample size of $1,500$), changing the main effects of the four biomarkers $\beta_{X_{21}}, \beta_{X_{41}}, \beta_{X_{61}}, \beta_{X_{81}}$.}
\label{fig:power}
\end{figure}

In Fig.~\ref{fig:power}(a), with highly correlated biomarkers, the proposed ridge regression screening procedure demonstrated substantially higher power than the univariate screening procedures, showing a clear benefit of accounting for correlations between the biomarkers at stage 1. For the univariate screening procedures, all the biomarkers with univariate marginal signals, including $X_1, \dots, X_{100}$, were likely to be retained after screening in the ``threshold'' approach or land into the top buckets at stage 2 in the ``rank'' approach. In contrast, the ridge screening procedure considered the effect of each biomarker, adjusted for all other biomarkers, and therefore tended to ascribe less evidence to biomarkers whose marginal associations were exaggerated by correlation with the true signal(s). Thus, biomarkers with true marginal associations, which are more likely to have interactions, tended to be ranked in the top buckets because of accounting for biomarker-biomarker correlations at stage 1. This enhanced the power of the overall two-stage approach compared with using the univariate screening procedures. In Fig.~\ref{fig:power}(b), with independent biomarkers, where the multivariate regression is not required for unbiased effect estimation, the univariate screening and the ridge screening procedures using weighted hypothesis tests perform similarly. All three two-stage tests outperformed the single-step interaction test by providing better power at the same family-wise error rate level whether biomarkers are correlated or independent.

In Fig.~\ref{fig:power}(c), we simulated scenarios with one biomarker having an interaction, no correlations among the biomarkers, and changed only the main effect of the interacting biomarker $\beta_{X_1}$, i.e. main effects of the other four biomarkers were the same as the previous scenario. The sample size was fixed at $1,500$. Fig.~\ref{fig:power}(c) reveals that there are some special cases, in which the main and interaction effect parameters are in opposite directions such that they cancel out, where all two-stage approaches give lower power than a single-step test.

In Fig.~\ref{fig:power}(d), we used the previous scenario with one biomarker having an interaction (biomarker correlation $\rho = 0.6$, sample size of $1,500$) as the base, and changed only the main effects of the four biomarkers with main effects alone $\beta_{X_{21}}, \beta_{X_{41}}, \beta_{X_{61}}, \beta_{X_{81}}$. Fig.~\ref{fig:power}(d) shows that power of all four tests decreases with increasing effect sizes of main-effect only biomarkers, because the proportion of variation explained by the interaction-effect biomarker decreases. The univariate screening using weighted hypothesis testing performs worse than the single-step test when effect sizes of four main-effect biomarkers become too large. This is because a large number of biomarkers that only have marginal associations, and no interaction, tend to fall into the top buckets, thus the bucket size allocated to the true interaction signal can lead to a more stringent significance threshold than that allocated by the single-step test through the Bonferroni adjustment accounting for all $m$ biomarkers. The ridge screening strategy still outperforms the single-step test, despite the biomarkers with marginal effects only exhibiting very strong stage 1 associations; their many correlated proxies are still screened out through multivariate modelling.

In Web Appendix E, we summarize family-wise error rates in different scenarios, which shows no inflation for all the screening procedures. We also provide additional simulation results. Relative patterns of performance among the screening strategies were consistent with the results described above, demonstrating the robustness of our method and findings.

\subsection{Data Applications}

In addition to validating our methods through simulations, we exemplified our approaches in two real data applications. 

We first applied our approaches to data from the randomized controlled trial START \citep{fonagy2020multisystemic}, which is composed of $684$ participants aged from $11$ to $17$ with antisocial behavior, half of whom were treated with management as usual (the control arm) and the rest were treated with multisystemic therapy followed by management as usual (the treatment arm). We used a secondary outcome of this trial, the $18$ months' follow-up outcome from Inventory of Callous and Unemotional Traits, as the continuous outcome and applied our interaction testing procedures to detect covariates having interactions with the treatment. We excluded covariates with more than $10\%$ missing data and used mean imputation to replace missing values for covariates with less than $10\%$ missing data. As a result, $75$ covariates were included in the analysis. Correlation among these covariates is generally low (a correlation plot is provided in Web Appendix F).

We performed all four screening procedures described in the previous section with a significance level of $\overline{\alpha} = 0.05$ and did not find any significant interactions. The top covariates from each of the univariate screening and ridge screening procedures are presented in Table~\ref{tb:biomarkersScreening}, which shows that the selected covariates from these two procedures are similar in this data set where covariates have low correlation.

In the second application, we applied our approaches retrospectively to a publicly available dataset with high-dimensional gene expression biomarkers (the PREVAIL trial) \citep{muscedere2018prevention}. The dataset is a phase II randomized trial which aimed to evaluate the efficacy of lactoferrin as a preventative measure for hospital-acquired infections. Gene expression data are available for $61$ patients from the National Center for Biotechnology Information (NCBI) website (GSE118657). Of the $61$ patients, $32$ patients were in the lactoferrin group, and the remaining patients were in the placebo group. We used the Sequential Organ Failure Assessment (SOFA) score measuring change in organ function post-randomization as the continuous response endpoint. From a total of $49,495$ genes, we restricted our analysis to the $10,000$ probes with the highest variability.

All four methods described in the previous section with a significance level of $\overline{\alpha} = 0.05$ did not find any significant biomarker-treatment interactions. A list of the top biomarkers from different marginal screening procedures is presented in Table~\ref{tb:biomarkersScreening}. The rankings of selected covariates are notably different between the ridge regression screening and the univariate screening procedures, likely owing to the high correlation among the biomarkers.

In addition, we examined the empirical correlation between stage 1 ridge screening and stage 2 interaction test statistics applied in the above two real data sets. Table~\ref{tb:correlation} summarizes results from Pearson correlation tests, which shows that the empirical correlation between stages is close to zero and in all cases the $95\%$ confidence interval contains zero as expected.
\begin{table}
	\def~{\hphantom{0}}
	\caption{Top covariates from different stage 1 marginal screening procedures}
	\begin{tabular}{|p{0.2cm}|p{7.5cm}|p{7.5cm}|}
		\Hline
		& \multicolumn{2}{c}{START trial} \\
		& Univariate screening & Ridge screening \\
		\hline
		1 & Total Inventory of Callous and Unemotional Traits & Total Inventory of Callous and Unemotional Traits \\
		2 & Total Antisocial Beliefs and Attitudes Scale & Total Antisocial Beliefs and Attitudes Scale \\
		3 & Strengths \& Difficulties Conduct Problems Score & Strengths \& Difficulties Conduct Problems Score \\
		4 & Strengths \& Difficulties ProSocial Behaviour Score & Strengths \& Difficulties ProSocial Behaviour Score \\
		5 & Strengths \& Difficulties Hyperactivity Score & Strengths \& Difficulties Hyperactivity Score \\
		6 & Volume of self reported delinquency excluding violence towards siblings & Volume of self reported delinquency excluding violence towards siblings \\
		7 & Strengths \& Difficulties Total Difficulties Score & Strengths \& Difficulties Total Difficulties Score \\
		8 & IQ & IQ \\
		9 & Variety of self reported delinquency excluding violence towards siblings & Parental reported total Inventory of Callous and Unemotional Traits \\
		10 & Parent reported Strengths \& Difficulties Conduct Problems Score & Alabama Positive Parental Involvement Score \\
		\Hline
		& \multicolumn{2}{c}{PREVAIL trial} \\
		& Univariate screening & Ridge screening \\
		\hline
		1 & 11715617\_a\_at & 11715488\_s\_at \\
		2 & 11749774\_x\_at & 11715489\_a\_at \\
		3 & 11725694\_at & 11739745\_a\_at \\
		4 & 11746124\_x\_at & 11749774\_x\_at \\
		5 & 11739745\_a\_at & 11746124\_x\_at \\
		6 & 11747047\_a\_at & 11747047\_a\_at \\
		7 & 11715488\_s\_at & 11728717\_at \\
		8 & 11720970\_at & 11725694\_at \\
		9 & 11751473\_a\_at & 11716479\_s\_at \\
		10 & 11756156\_s\_at & 11752423\_a\_at \\
		\hline
	\end{tabular}
	\label{tb:biomarkersScreening}
\end{table}
\begin{table}
	\def~{\hphantom{0}}
	\caption{Empirical correlation between stage 1 ridge screening and stage 2 interaction test statistics}
	\begin{tabular}{lcc}
		\Hline
		& START & PREVAIL \\
		\hline
		Estimate & 0.044 & 0.001 \\
		$p$-value & 0.711 & 0.938 \\
		95\% confidence interval & (-0.188, 0.271) & (-0.019, 0.020) \\
		\hline
	\end{tabular}
	\label{tb:correlation}
\end{table}

\section{Discussion}

We propose, for the first time with formal justification, the use of ridge regression in a two-stage interaction testing framework for identifying biomarker signatures of treatment efficacy in randomized clinical trials. Interaction testing frameworks which are designed to scale to large numbers of covariates will become ever more important as -omics technologies continue to drop in price and become routinely measured in clinical trials. Naturally, there will be variation in the level of correlation among different sets of -omics biomarkers from one setting to the next. For instance, when there is a strong apriori hypothesis of which genes influence treatment efficacy, such that a panel of genetic markers are all taken from the same region, pairwise correlations will be stronger on average compared to a genome-wide panel of variants, because local genetic correlations tend to be much stronger than long-range correlations (known as linkage disequilibrium decay). Similarly, considering transcriptomics, correlations will be stronger when focusing on a subset of genes that correspond to the same pathway. Therefore the ridge screening approach will be particularly well motivated when related biomarkers of apriori interest have been pre-selected, for instance from a gene region or pathway. These biomarker sets will tend to exhibit the strongest correlation structures, and so will benefit the most from multivariate modeling during stage 1 screening.

It is known that ridge regression has a tendency to average effects across strongly correlated covariates. This phenomenon is not desirable for a screening strategy since it could inflate the number of non-interacting biomarkers being put forward to stage 2. Thus, lasso \citep{tibshirani1996regression}, as an alternative penalized regression model which does not exhibit this effect-averaging behavior, may be expected to perform better. However, as lasso uses a $L_1$ penalty which is not a smooth function, it is challenging to prove it meets the between-stage independence requirement to preserve the overall family-wise error rate in two-stage approaches. Since the main goal of employing the penalized regression screening procedures in stage 1 is to account for biomarker-biomarker correlations, some less computationally intensive multiple testing correction methods for correlated tests might be beneficial \citep{nyholt2004simple, gao2008multiple}. However, applying such methods which calculate an ``effective'' number of independent tests to the single-step interaction test in a limited set of simulations did not offer any power improvement when controlling for the same family-wise error rate (results not shown). We suggest further investigation in how to incorporate these methods into the two-stage interaction framework including a formal justification of the family-wise error rate control as a topic of future work.

We also showed that there exist special cases where our proposed two-stage screening strategy offers no benefit, e.g. the case when the main effect of a biomarker and its interaction effect with the treatment to the response are in opposite directions, which reduces the strength of the marginal association (sometimes leaving no detectable marginal effect) for true interactions. We suggest exploring the weighting scheme thus changing how much stage 1 information to be used in the following stage 2 tests as a future topic for investigation. Another technical caveat was shown by \citet{sun2018testing} that, for logistic regression, the interaction estimator under treatment misspecification can be biased when the biomarker is associated either indirectly or directly with the outcome. This is a generic issue to interaction modeling using logistic regression, but could manifest in our framework as an elevated family-wise error rate at stage 2 one-biomarker-a-time tests. Therefore, we highlight that, currently, our theoretical work only guarantees family-wise error rate control when using linear regression. The extent to which this bias might inflate family-wise error rates when applying our framework using logistic regression, and potential corrections, will be the topic of future work.


\backmatter


\section*{Acknowledgements}

This work was funded by the UK Medical Research Council (grant number MR/R502303/1 to J.W., grant number MC\_UU\_00002/9 to A.P. and P.J.N., grant number MC\_UU\_00002/6 to J.M.S.W.). P.J.N. acknowledges support from the NIHR Cambridge Biomedical Research Centre. The authors thank the START trial investigators for use of their data.

\section*{Data Availability Statement}

START data can be accessed through the procedure described in \citet{fonagy2020multisystemic}. PREVAIL data were derived from the NCBI website (\url{https://www.ncbi.nlm.nih.gov/geo/query/acc.cgi?acc=GSE118657}) \citep{maslove2018time}.

\vspace*{-8pt}


%
\bibliographystyle{biom}
\bibliography{../../references}


\section*{Supporting Information}

Web Appendices referenced in Sections~\ref{sec:methods} and \ref{sec:res}, and the R code for simulation studies and data applications are available with this paper at the Biometrics website on Wiley Online Library.

\vspace*{-8pt}

\appendix


\section{}
\subsection{Proof of Theorem 1}

Based on the unified approach to proving between-stage asymptotic independence by \citet{dai2012two}, we need to evaluate the covariance matrix $\boldsymbol{A}_1^{-1} \boldsymbol{B} \boldsymbol{A}_2^{-1}$, where 
\begin{eqnarray}
&& \boldsymbol{A}_1 = E[(\boldsymbol{X}_i \boldsymbol{X}_i^T) \{Y_i - E(Y_i \mid \boldsymbol{X}_i)\}^2] \nonumber \\
&& \boldsymbol{B} = E[(\boldsymbol{X}_i \boldsymbol{V}_{ij}^T) \{Y_i - E(Y_i \mid \boldsymbol{X}_i)\} \{Y_i - E(Y_i \mid \boldsymbol{V}_{ij})\}] \nonumber \\
&& \boldsymbol{A}_2 = E[(\boldsymbol{V}_{ij} \boldsymbol{V}_{ij}^T) \{Y_i - E(Y_i \mid \boldsymbol{V}_{ij})\}^2] \nonumber
\end{eqnarray}

We simplify the expression of $\boldsymbol{B}$ as 
\begin{eqnarray}
\boldsymbol{B}
&=& E[(\boldsymbol{X}_i \boldsymbol{V}_{ij}^T) \{Y_i^2 - Y_i E(Y_i \mid \boldsymbol{X}_i) - Y_i E(Y_i \mid \boldsymbol{V}_{ij}) + E(Y_i \mid \boldsymbol{X}_i) E(Y_i \mid \boldsymbol{V}_{ij})\}] \nonumber \\
&=& E[(\boldsymbol{X}_i \boldsymbol{V}_{ij}^T) E\{Y_i^2 - Y_i E(Y_i \mid \boldsymbol{X}_i) - Y_i E(Y_i \mid \boldsymbol{V}_{ij}) + E(Y_i \mid \boldsymbol{X}_i) E(Y_i \mid \boldsymbol{V}_{ij}) \mid \boldsymbol{X}_i\}] \nonumber \\
&=& E(\boldsymbol{X}_i \boldsymbol{V}_{ij}^T) var(Y_i \mid \boldsymbol{X}_i) \nonumber
\end{eqnarray}
which uses the law of iterated expectations, the fact that $\boldsymbol{X}_i$ includes $\boldsymbol{V}_{ij}$ under the null hypothesis $\beta_{X_j \times T} = 0$, and assumes homogeneity of variance, i.e. $var(Y_i \mid \boldsymbol{X}_i)$ is a constant.

Similarly, we have $\boldsymbol{A}_1 = E(\boldsymbol{X}_i \boldsymbol{X}_i^T) var(Y_i \mid \boldsymbol{X}_i)$ and $\boldsymbol{A}_2 = E(\boldsymbol{V}_{ij} \boldsymbol{V}_{ij}^T) var(Y_i \mid \boldsymbol{V}_{ij})$. Thus, 
\begin{eqnarray}
\boldsymbol{A}_1^{-1} \boldsymbol{B} \boldsymbol{A}_2^{-1} \propto E(\boldsymbol{X}_i \boldsymbol{X}_i^T)^{-1} E(\boldsymbol{X}_i \boldsymbol{V}_{ij}^T) E(\boldsymbol{V}_{ij} \boldsymbol{V}_{ij}^T)^{-1} \nonumber
\end{eqnarray}
We consider the second and the third terms 
\begin{eqnarray}
\underset{(m + 1) \times 3}{E(\boldsymbol{X}_i \boldsymbol{V}_{ij}^T)}
&=&
\left\{\begin{array}{ccc}
E(T_i X_{ij}) & E(T_i^2) & E(T_i^2 X_{ij}) \\
E(X_{i1} X_{ij}) & E(T_i X_{i1}) & E(T_i X_{i1} X_{ij}) \\
\vdots & \vdots & \vdots \\
E(X_{im} X_{ij}) & E(T_i X_{im}) & E(T_i X_{im} X_{ij})
\end{array}\right\} \nonumber \\
\underset{3 \times 3}{E(\boldsymbol{V}_{ij} \boldsymbol{V}_{ij}^T)^{-1}}
&=& \frac{1}{det\{E(\boldsymbol{V}_{ij} \boldsymbol{V}_{ij}^T)\}}
\left\{\begin{array}{ccc}
\cdot & \cdot & E(T_i X_{ij}) E(T_i^2 X_{ij}) - E(T_i^2) E(T_i X_{ij}^2) \\
\cdot & \cdot & E(T_i X_{ij}) E(T_i X_{ij}^2) - E(X_{ij}^2) E(T_i^2 X_{ij}) \\
\cdot & \cdot & E(X_{ij}^2) E(T_i^2) - E(T_i X_{ij})^2
\end{array}\right\} \nonumber
\end{eqnarray}

Thus, for the $(m + 1) \times 3$ matrix $E(\boldsymbol{X}_i \boldsymbol{V}_{ij}^T) E(\boldsymbol{V}_{ij} \boldsymbol{V}_{ij}^T)^{-1}$, the $(k + 1)$th element ($k = 1, \dots, m$) of the last column is proportional to 
\begin{eqnarray}
&& \left\{\begin{array}{ccc}
E(X_{ik} X_{ij}), & E(T_i X_{ik}), & E(T_i X_{ik} X_{ij})
\end{array}\right\}
\cdot \left\{\begin{array}{c}
E(T_i X_{ij}) E(T_i^2 X_{ij}) - E(T_i^2) E(T_i X_{ij}^2) \\
E(T_i X_{ij}) E(T_i X_{ij}^2) - E(X_{ij}^2) E(T_i^2 X_{ij}) \\
E(X_{ij}^2) E(T_i^2) - E(T_i X_{ij})^2
\end{array}\right\} \nonumber \\
&=& E(T_i) var(T_i) E(X_{ij}) \{E(X_{ik} X_{ij}) E(X_{ij}) - E(X_{ik}) E(X_{ij}^2)\} = 0 \nonumber
\end{eqnarray}
which uses the independence between $T_i$ and $X_{ij}$, and the assumption $E(T_i) = 0$ or $E(X_{ij}) = 0$. Similarly, the first element of the last column is also zero.

Premultiplying $E(\boldsymbol{X}_i \boldsymbol{V}_{ij}^T) E(\boldsymbol{V}_{ij} \boldsymbol{V}_{ij}^T)^{-1}$ by $E(\boldsymbol{X}_i \boldsymbol{X}_i^T)^{-1}$ completes the covariance matrix, the last column of which are all zeros. Thus, for any $j = 1, \dots, m$, we have $cov\{n^{1 / 2} (\widehat{\delta}_{X_j}^0 - \delta_{X_j}), n^{1 / 2} (\widehat{\beta}_{X_j \times T} - \beta_{X_j \times T})\} \rightarrow 0$ in probability.

\label{lastpage}

\end{document}